\documentclass[aps,prd,preprint,superscriptaddress,showpacs]{revtex4}
\usepackage{slashed}
\usepackage{graphicx}
\usepackage{verbatim}
\usepackage{ulem}
\usepackage{color}
\definecolor{My_red}        {cmyk}{0.00,1.00,1.00,0.20}

\newcommand{\bmat}{\left(\begin{array}}
\newcommand{\emat}{\end{array}\right)}
\newcommand{\beq}{\begin{equation}}
\newcommand{\eeq}{\end{equation}}
\newcommand{\wt}{\widetilde}

\def\ra{\rightarrow}

\def\Ld{\Lambda}
\def\ld{\lambda}
\def\f{\frac}

\def\bwt{\begin{widetext}}
\def\ewt{\end{widetext}}
\def\be{\begin{equation}}
\def\ee{\end{equation}}
\def\bea{\begin{eqnarray}}
\def\eea{\end{eqnarray}}
\def\bean{\begin{eqnarray*}}
\def\eean{\end{eqnarray*}}
\def\bary{\begin{array}}
\def\eary{\end{array}}
\def\bit{\begin{itemize}}
\def\eit{\end{itemize}}

\def\ra{\rightarrow}

\def\Ld{\Lambda}

\def\ld{\lambda}

\def\su5u1{SU(5) \times U(1)}
\def\fsu5u1{SU(5) \times U(1)'}
\def\so10{SO(10)}
\def\sq20{SO(10) \times SO(10)}

\def\ra{\rightarrow}

\def\Ld{\Lambda}
\def\ld{\lambda}
\def\f{\frac}

\def\L{\left(}
\def\R{\right)}

\def\ra{\rightarrow}

\def\Ld{\Lambda}

\def\ld{\lambda}

\def\su5u1{SU(5) \times U(1)}
\def\fsu5u1{SU(5) \times U(1)'}
\def\so10{SO(10)}
\def\sq20{SO(10) \times SO(10)}

\usepackage[centertags]{amsmath}
\usepackage{amssymb}

\begin{document}

\title{ Natural $X$-ray Lines from the Low Scale Supersymmetry Breaking }

\author{Zhaofeng Kang}
\email[E-mail: ]{zhaofengkang@gmail.com}
\affiliation{Center for High-Energy Physics, Peking University, Beijing, 100871, P. R. China}
\affiliation{School of Physics, Korea Institute for Advanced Study,
Seoul 130-722, Korea}

\author{P. Ko}
\email[E-mail: ]{pko@kias.re.kr}
\affiliation{School of Physics, Korea Institute for Advanced Study,
Seoul 130-722, Korea}

\author{Tianjun Li}
\email{tli@itp.ac.cn}

\affiliation{State Key Laboratory of Theoretical Physics
and Kavli Institute for Theoretical Physics China (KITPC),
Institute of Theoretical Physics, Chinese Academy of Sciences,
Beijing 100190, P. R. China}

\affiliation{School of Physical Electronics,
University of Electronic Science and Technology of China,
Chengdu 610054, P. R. China}

\author{Yandong Liu}
\email{ydliu@itp.ac.cn}

\affiliation{State Key Laboratory of Theoretical Physics
and Kavli Institute for Theoretical Physics China (KITPC),
Institute of Theoretical Physics, Chinese Academy of Sciences,
Beijing 100190, P. R. China}

\date{\today}

\begin{abstract}

In the supersymmetric models with low scale supersymmetry (SUSY) breaking where the gravitino mass is around keV,  we show that the 3.5 keV $X-$ray lines can be explained naturally through several different mechanisms: (I) A keV scale dark gaugino plays the role of sterile neutrino in the presence of bilinear $R-$parity violation. Because the light dark gaugino obtains Majorana mass only via gravity mediation, it is a decaying warm dark matter (DM) candidate; (II) The compressed cold DM states, whose mass degeneracy is broken by gavity mediated SUSY breaking, emit such a line via the heavier one decay into the lighter one plus photon(s). A highly supersymmetric dark sector may readily provide such kind of system; (III) The light axino, whose  mass again is around the gravitino mass, decays to neutrino plus gamma in the $R-$parity violating SUSY. Moreover, we comment on dark radiation from dark gaugino.

\end{abstract}

\maketitle

\section{Introduction and motivations}

Dark matter (DM) as a solid evidence for new physics beyond the standard model (SM) receives wide attention,
both from the theoretical and experimental physics communities. The original focus is the weakly interacting massive
particle (WIMP) DM candidate due to the WIMP miracle argument on the correct order
of DM relic density. But the experimental results discouraged us despite of several events which are far
from confirmation. On the other hand, the keV scale warm DM (WDM) become more interesting
since the progress on $N-$body simulation of DM structure formation indicates that WDM
can give the correct abundance of DM galaxy substructures
nevertheless cold DM can not~\cite{subs}.

If DM is in the keV region, the searching strategy will be quite different. For instance, it, which is non-relativistic in the present epoch, tends to leave null hints in the current underground DM detectors (This seems to be consistent with the most stringent bounds from the XENON100~\cite{XENON100} and LUX~\cite{LUX} experiments.). On the top of that, its signatures from the sky should not locate at the high energy region, and we may have to rely on the observation of $X-$ray line, which has relatively clear astrophysical background.  Recently, a tentative 3.5 keV gamma ray line through the observation of galaxy clusters and the Andromeda galaxy was discovered~\cite{Boyarsky:2014jta}. Although the origin of this line is controversial~\cite{Jeltema:2014qfa,Boyarsky:2014paa},  it is still of interest to ascribe it to dark matter activities.


This possibility inspires a lot of works soon later. Refs.~\cite{Krall:2014dba,Frandsen:2014lfa} attempt to understand it via the effective operator analyses. Specific candidates are also proposed, e.g., a keV sterile neutrino~\cite{Abazajian:2014gza} which was motivated long ago, axion-like particles~\cite{Jaeckel:2014qea} (DM decays into axion which then converts into a pair of photons may fit data better~\cite{Cicoli:2014bfa}), axino~\cite{Kong:2014gea}, and millicharged dark matter~\cite{Aisati:2014nda}. All these particles have masses at the keV scale, producing the $X-$ray line via decaying or annihilating into gamma among others. But the eXciting DM (eXDM)~\cite{Finkbeiner:2014sja} takes a quite different approach which is beyond the WDM framework. There the $X-$ray line instead comes from the decay of the heavy DM exciting state, which is tinily heavier than DM by an amount about 3.5 keV, back into the DM plus photon with others.

We should seriously ponder on the natural origin of the keV scale, which is by no means a trivial question
in model building since the SM has a characteristic scale 100 GeV. In the keV WDM scenario, one may want to seek a theoretical reason for such a low mass scale. In the eXDM-like scenario, generating such a small mass splitting without incurring fine-tuning is of concern to us as well. It is well-known that supersymmetry provides a natural solution to the gauge hiearchy problem. And in the supersymmetric SMs (SSMs) with $R$-parity, we can realize gauge coupling unification and have a dark matter
candidate. In particular, if SUSY is broken around $10^6$ GeV, gravitino mass is indeed around keV.  Thus,
 both keV scale WDM and keV scale mass splitting can be tied to the keV gravitino mass if they
are generated from gravity mediated SUSY breaking. In this paper, three examples are presented to show how they are related, and of course how their decay final states include  gamma: A light dark gaugino or axino, obtaining a Majorana mass $\sim m_{\wt G}$ via gravity mediated SUSY breaking, can decay into a neutrino plus photon in the presence of bilinear $R-$parity violation; A highly supersymmetric dark sector with SUSY-breaking $\sim m_{\wt G}$ provides an eXDM-like system (or compressed DM system dubbed in our paper), with mass splitting between DM states set by $m_{\wt G}$. The heavier DM state three-body decay into a pair of gamma and a lighter DM state via a light CP-odd particle exchange. Furthermore, we may explain dark radiation via dark gaugino.

This paper is organized as follows. In Section II we propose that in the SSMs with bilinear $R-$parity violation
and low scale SUSY-breaking, a light dark gaugino can naturally be a sterile neutrino. In Section III we explore the idea of $X-$ray line from a compressed WIMP-like system, which is also naturally accommodated in a highly supersymmetric DM sector receiving a small SUSY-breaking via gravity mediation. In Section IV we briefly study the connection between axino mass and low scale SUSY-breaking. Our remarks on dark radiation is in Section V. And
discussion and conclusion are casted in Section IV. Some complementary details are given in the Appendix.

\section{The Dark Gaugino Models}

In the SM extended with canonical seesaw mechanism, the right-handed neutrino at keV scale is a well motivated
WDM candidate~\cite{WDM:st}. But a simple understanding of such a low seesaw scale, to our knowledge, is absent. In this section we will show that in the supersymmetric SMs with a dark gauge group $U(1)_X$, the dark gaugino can naturally be a keV sterile neutrino.

\subsection{Supersymmetric $U(1)_X$ model with kinetic mixing}

A local Abelian symmetry $U(1)_X$ is extensively studied in the dark matter models with various motivations~\cite{U(1)X,Kmixing,Kang:2010mh}. In this article we are interested in its supersymmetric version, focusing on the kinetic part
\begin{align}\label{}
\f{1}{4}\int d^2\theta\L W_YW_Y+W_XW_X-2\epsilon W_YW_X\R+c.c.,
  \end{align}
where $W_{Y,X}$ are the vector superfields of $U(1)_{Y,X}$ gauge fields respectively. The last term is the kinetic mixing term which connects the dark and visible sectors but is suppressed by $\epsilon\ll1$. The mass spectrum in the complete $U(1)_X$ model is model dependent, and either light or heavy is acceptable in our setup, although the mass spectrum around keV may be more natural (In appendix~\ref{split} we give an example to show how a heavy dark spectrum leaves 
a light dark gaugino.). We only require that the dark gaugino $\wt X$ be as light as a few keVs. This mission
is possible since the gaugino mass is a result of gravity mediated SUSY breaking, and thus can be separated from other scales in the $U(1)_X$ sector, e.g., the dark vector boson mass. But we have to ensure that dark gaugino can not fast decay into gravitino plus dark states like the dark vector boson, which can be forbidden kinematically. The detailed model building is out of the scope of this paper.

\begin{figure}[htb]
\includegraphics[width=5.2in]{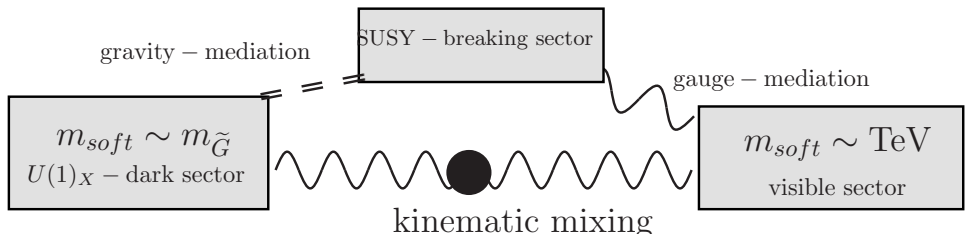}
   \caption{A schematic diagram shows the SUSY-breaking mediation scheme which generates the large SUSY breaking soft mass hierarchy among the dark and visible sectors. One can find a similar but more complicated diagram in Ref.~\cite{Kang:2010mh}, which attempts to generate the GeV scale dark sector.
   \label{darkU} }
\end{figure}

We now proceed to show how SUSY-breaking mediation generates a large soft mass hierarchy between the $U(1)_X$ sector and the visible sector. The idea is simple, see a schematic diagram of the dynamics in Fig.~\ref{darkU}. SUSY is first broken in the hidden sector at a low scale $\sqrt{F}\ll M_{\rm Pl}$. Via suitable mechanisms such as 
gauge mediated SUSY-breaking (GMSB), SUSY-breaking is then transmitted to the visible sector, generating the phenomenologically desired TeV-scale soft spectrum. But at this step the dark sector does not feel SUSY-breaking and in particular the dark gaugino is still massless. The universal gravity mediation contributes to dark gaugino mass, estimated to be
\begin{align}\label{}
m_{\wt X} \sim {\cal O} (m_{\wt G})\simeq {\cal O}(F/\sqrt{3}M_{\rm pl})~,
  \end{align}
up to an order one coefficient. Therefore, $\wt X$ is indeed at the keV scale if the SUSY-breaking scale $\sqrt{F}$ is as low as 1000 TeV. As one can see, our setup for separating SUSY breaking soft mass scales can be irrelevant to dark gauge symmetry breaking, so it applies to a large class of $U(1)_X$ models.

By virtue of its lightness, the potential final states of $\wt X$ decay are restricted to $\wt G$, neutrino $\nu$, photon $\gamma$ and gluon $g$. We should first figure out the relevant interactions. For that purpose, it is convenient to employ field shifts for the gauge fields~\cite{Kmixing}: $A_\mu'= A_\mu -\epsilon \cos\theta_w X_\mu$ and $X_\mu'= X_\mu +\epsilon \sin\theta_w Z_\mu$. There $A_\mu$ and $Z_\mu$ are respectively the massless $U(1)_{\rm EM}$ gauge field and massive $Z-$boson after eletroweak symmetry breaking. Similarly, the dark gaugino is shifted by the hypercharge gaugino field
\begin{align}\label{}
\wt X'=\wt X+\epsilon \wt B~.
  \end{align}
After these shifts, all of the kinetic mixings are eliminated up to ${\cal O}(\epsilon^3)$. In this basis, the leading order interaction between dark gaugino and visible sectors matters are encoded in
\begin{align}\label{}
\epsilon {\cal O}(m_{\wt X}/m_{\wt B}){\wt X}\wt J_B+c.c.,{\rm\,\,with\,\,\,} \wt J_B=g_Y\sum_{f}Q_f\wt f^\dagger f,
  \end{align}
with $Q_f$ the $U(1)_Y$ charge of $f$. In particular, the effective vertex $\wt X \wt \nu_L \nu_L$ is generated at this order.

\subsection{The bilinear $R-$parity violation}

To open a channel for dark gaugino decay into gamma, we further introduce the bilinear $R-$parity violating terms
\begin{align}\label{BRP}
W_{R}=\mu_iL_iH_u~,
  \end{align}
which induces the non-vanishing Vacuum Expectation Values (VEVs) for sneutrinos $\wt\nu_L$ (For simplicity, only one flavor of sneutrino will be considered hereafter.). Its concrete value depends on the SUSY breaking soft parameters and the Higgs parameters, so we do not discuss the details here. The constraints from its contribution to the tree-level neutrino mass are mild and the GeV scale VEV is allowed~\cite{Barbier:2004ez}. By virtue of the sneutrino VEV, now there is a mixing between $\wt X$ and $\nu_L$. Assuming that the mixing term is much smaller than the dark gaugino mass, then the mixing angle is given by
\begin{align}\label{}
\sin2\theta\simeq 2\epsilon\f{g_Y}{2}\f{m_{\wt X}}{m_{\wt B}}\f{\langle{\wt \nu}_L\rangle}{m_{\wt X}}=\epsilon g_Y \f{\langle{\wt \nu}_L\rangle}{m_{\wt B}},
  \end{align}
independent on the dark gaugino mass.

In reality, the dark gaugino now becomes nothing but a keV scale sterile neutrino, and therefore it radiatively decays into an active neutrino plus an photon via the $W-$loop. The decay width
is well known~\cite{WDM:st} as follows
\begin{align}\label{WDM:G}
\Gamma_{\nu\gamma}\simeq 1.62\times 10^{-28}s^{-1}\L\f{\sin^22\theta}{7\times10^{-11}}\R\L\f{m_{\wt X}}{\rm7keV}\R^5~.
  \end{align}
We have parametrized the required decay rate for 7~keV WDM to fit the data~\cite{Boyarsky:2014jta}. Accommodating $\sin2\theta\sim10^{-5}$ is not difficult provided that $\epsilon$ and the sneutrino VEV are not too small. Typically, one needs $\epsilon\sim {\cal O}(10^{-2})$ and $\langle{\wt \nu}_L\rangle\sim{\cal O}(\rm GeV)$ for a weak scale bino. Note that the resulted contribution to active neutrino mass $m_\nu\sim 10^{-7}$ eV is too small to account for the active neutrino masses and will not affect the neutrino masses and mixings. But if dark matter only contains a small fraction of dark gaugino, say the gravitino is the dominant one, then the mixing angle with neutrino(s) thereof is allowed to be large enough. We actually do not have to insist on that dark gaugino should produce the correct neutrino phenomenologies, since with $\mu_iL_iH_u$ the visible sector can~\cite{Barbier:2004ez}.

Several comments are in order. Firstly, dark gaugino sterile neutrino has a remarkable difference to the conventional sterile neutrino, i.e., its interactions are not specified by the mixing with active neutrino. At the leading order of $\epsilon$, it participates in gauge interactions with the dark sector particles which are not specified here and interacts with gravitino, which are enhanced by low SUSY breaking scale. As a consequence it has a normal thermal history (the dark sector establishes thermal equilibrium with the visible sector via kinetic mixing). Secondly, the light gravitino, with decay rate suppressed by $1/M_{\rm Pl}^4$, is still sufficiently long lived thus being a member of DM today (As a matter of fact, the keV gravitino also decays into gamma plus neutrino, but the decay width is extremely small thus negligible except for~\cite{Bomark:2014yja}.). In our paper both $\wt X$ and $\wt G$ are keV scale warm DMs, and they have comparable relic density if they decouple roughly in the same epoch, say around 100 GeV as follows
\begin{align}\label{DM-Density}
\Omega h^2\simeq 0.11\L\f{m_{\wt X.\wt G}}{\rm 10 ~keV}\R \L\f{106.75}{g_{*S,f}}\R \L\f{100}{{\cal S}}\R,
  \end{align}
with $g_{*S,f}$ the relativistic degrees of freedom at $\wt G$ decoupling. So, we may need an entropy release factor ${\cal S}\sim$100 to dilute them, and interestingly they may be furnished by the $U(1)_X$ sector. We leave this for further study.

\section{The Compressed WIMP Models}

As stressed in the Introduction, although WDM is well motivated to explain the 3.5 keV line,
 it is of importance to examine whether or not the line can be naturally produced in the WIMP-like DM
since it  services as the most popular DM candidate. Its mass, broadly, varies from the GeV to TeV scale. Thus, gamma-ray line from that WIMP should locate at the higher energy spectrum. However, there is a good exception.
An illustrative example can be found in collider physics. To get around the stringent LHC bounds on scalar top quark $\wt t$, it was proposed that $\wt t$ has a close mass with the lightest supersymmetric particle (LSP) $\chi$ such that the produced visible particles from $\wt t\ra \chi+...$ are too soft to be detected. In this way, the compressed SUSY hides stop at
the LHC. What we learn from this is the lesson that heavy particle decay can produce a soft spectrum, if the decay proceeds near the kinematic threshold. However, we need to explain why the mass splitting is so small.


Now, we apply this idea to the WIMP-like models producing $X-$ray line. It is simply realized by replacing stop with the WIMP-like DM $X_1$ and the LSP with another DM state $X_2$, which is slightly lighter than $X_1$: $m_{X_1}-m_{X_2}=\delta\sim {\cal O}(\rm\,keV)$. Considering $X_1$ decay into $X_2$ plus the SM states, the kinematically allowed modes are neutrino/photon/gluon, depending on the mediator $M$ which connects the dark and SM sectors. Actually, we already have such an example, the eXciting DM (XDM) model~\cite{eDM}, in which the TeV scale DM $\chi$ has an exciting state $\chi^*$ and they have a small mass splitting $\delta\simeq $ MeV. The original idea is to transfer the DM kinematic energy to the $e^++e^-$ pair via the up scattering process $\chi\chi\ra \chi^*\chi^*$ followed by the promptly de-exciting process $\chi^*\ra\chi+e^+e^-$. But in the case of keV mass splitting, the heavier state is expected to be so stable that it is a component of DM today. In this Section we follow this line~\footnote{In the preparation of this paper, several papers also considered this possibility~\cite{Frandsen:2014lfa,Allahverdi:2014dqa}. They studied two-body decay while we study three-body decay. The produced photon spectrum here is not exactly monochromatic, but the width is checked to be fairly narrow and thus it is supposed to be acceptable. More precise prediction requires detailed data fitting, which is beyond the scope of this paper.}, and propose that the compressed WIMP-like DM system is also able to produce $X-$ray line. In particular, such a system has a natural realization in the low scale SUSY breaking models. We will first make some model independent analyses, which will give some helpful observations. And then the working models will be presented.


\subsection{Preliminary model building}



A simple realization of the above compressed WMIP system is a nearly supersymmetric sector which, minimally, consists of a chiral superfield $X$.
\begin{align}\label{}
W=&\f{M_X}{2}X^2,\cr
{\cal L}_{\rm soft}=&c_1m_{\wt G}^2|X|^2+\L c_2\f{m_{\wt G}M_X}{2}X^2+c.c.\R ~,
  \end{align}
where $X$ is odd under a $Z_2$ symmetry and provides the compressed DM candidates. For later convenience, we decompose 
$X$ in terms of its real scalar components as $X=\f{1}{\sqrt{2}}\L X_R+iX_I\R$. It is not difficult to get the mass spectrum of the three dark states $X_{R/I}$ and $\wt X$
\begin{align}\label{spectrum}
m_{\wt X}=M_X,\quad m_{ X_{R/I}}\approx M_X\L1+\f{c_1}{2}\f{m_{\wt G}^2}{M_X^2}\pm \f{c_2}{2}\f{m_{\wt G}}{{M_X}}\R.
  \end{align}
Obviously, the bilinear soft terms dominantly account for mass splittings between dark states $X_{R/I}$ and $\wt X$, 
which is at the order of gravitino mass. Among them, the lightest one, assumed to be $X_I$ without loss of generality, is absolutely stable. The Majorana fermion $\wt X$ is additionally odd under the ordinary $R-$parity (assumed to be conserved), so it is also stable given that its mass splitting with $X_I$ is so small that $\wt X\ra X_I+\wt G$ is forbidden. In a similar way, the heaviest one $X_R$, with the previous assumptions, is stable as well.

We would like to  comment on the mechanism used here to split $X_{R/I}$. Concerning the superpotential alone,
the scalar potential of dark states possesses an enlarged $U(1)$ symmetry due to the holomorphic superpotential. This symmetry guarantees mass degeneracy between the CP-odd and -even components of $X$. However, SUSY-breaking, which is encoded in the soft terms, spoils holomorphy thus the accidental $U(1)$, generating small mass splitting at a scale set by gravitino mass. Later we will see that even in a realistic model where additional coupling to $X$ is introduced, this mechanism still works given that no appreciably SUSY breaking arises.

Before constructing a concrete model to address the $X-$ray line from $X_R$ decay, it is helpful to employ effective field theory analyses. The interactions between dark sector and visible sector need a portal which mediates $X_R$ decay into $X_I$ plus a pair of photons. In the case of CP-conservation, a CP-odd scalar, denoted as $a$ with a familiar axion effective coupling $a F\wt F$, is a good candidate. Thus, the effective Lagrangian is
\begin{align}\label{decay}
{\cal L}_{decay}=\mu_a aX_RX_I+\f{\alpha}{4\pi}\f{1}{8\Ld}a F_{\mu\nu}\wt F^{\mu\nu}.
  \end{align}
$X_R$ three-body decays into $X_I$ plus a pair of photons, mediated by $a$, with the following decay rate  
(see Appendix~\ref{three-body} for details)
\begin{align}\label{rate}
\Gamma_{eff}\simeq& \mu_a^2\f{\alpha^2}{107520\pi^5}
\f{\delta^7}{m_{X_R}^2m_a^4\Ld^2}\L \f{\delta}{m_{X_R}}\R\cr
=&1.14\times 10^{-28}s^{-1}\L\f{\mu_a}{1\rm\,GeV}\R^2\L
\f{\delta}{7.0\rm\,keV}\R^8\L\f{1.0\rm\,GeV}{m_{X_R}}\R^3
\L\f{40\rm\,GeV}{\Ld}\R^2\L\f{0.1\rm\,GeV}{m_a}\R^4.
  \end{align}
In the above expression of $\Gamma_{eff}$, we have multiplied by an extra factor $\delta/m_{X_R}$ by hand (thus it is labeled with ``eff"), taking into account the fact that the ratio of number densities of cold DM and WDM is $m_{\rm WDM}/m_{\rm CDM}$. With it, the parameterized value of $\Gamma_{eff}$ is still close to the one used in Eq.~(\ref{WDM:G}). We take $\delta=7.0$ keV instead of 3.5 keV because a pair of photon is produced.

Useful information on the three mass scales in Eq.~(\ref{rate}) are available, which guides the model building. The effective decay rate is greatly suppressed by $\delta^8$, so sufficiently light scales, or at least one of them, are needed to make it large enough. In particular, $a$ being a pseudo goldstone boson (PGSB)-like particle with mass below the GeV scale is of interest. The suppression scale $\Ld$ can have two kinds of origins: (I) The operator $aF\wt F$ is generated via a charged loop, and then $\Ld\simeq m_C/(4\sqrt{2} h_{aCC}N_CQ_C^2)$ with $(Q_C,\,m_C,\,h_{aCC})$ the (electronic charge, mass, coupling with $a$) of particle $C$ running in the loop~\cite{Kang:2012bq} and $N_C$ a color factor. Note that $\Ld$ may be much lighter than the estimated value, given a larger $Q_C$, say 2; (II) If the operator is generated by anomaly, $\Ld$ is associated with the spontaneously breaking of $U(1)_A$, which is quite model dependent.

\subsection{ Model building}

Now we will construct a concrete model which realizes the above idea. Asides from the dark vector field $X$,
the minimal model contains a singlet $S$ and a pair of charged vector-like particles $(C,\bar C)$. Its superpotential
 takes a form of
\begin{align}\label{portal}
W=&\f{M_X}{2}X^2+\f{\ld_X}{2}X^2S+\f{M_S}{2}S^2+\ld_C SC\bar C+m_CC\bar C.
  \end{align}
For simplicity, we have considered the model by dropping the irrelevant terms such as the cubic and linear terms of $S$. Adding them will not change our main discussions. $S$ does not acquire a VEV and therefore does not contribute to mass splitting between $X_{R}$ and $X_I$. The mass spectrum in the dark sector is exactly given by Eq.~(\ref{spectrum})~\footnote{Renormalization group flows induce a soft term for $M_XX^2/2$, but it is a three-loop effect, thereby suppressed greatly and negligible.}. As for the singlet $S$, for our purpose,
 all of its component masses can be commonly approximated to be $M_S$. In light of the previous analysis, it is favored to lie at a low scale, say around the sub-GeV scale.

At first, it is not difficult to derive the effective coupling $\mu_a$ (through the $F-$term of $S$) and effective suppression scale $\Ld$ introduced in Eq.~(\ref{decay}). They are respectively given by
\begin{align}\label{}
\mu_a&=-\ld_X M_S,\\
\Ld&=\f{1}{4\ld_C }\f{m_C}{N_CQ_C^2}.
  \end{align}
Note that the massive coupling $\mu_a$ is related to $M_S=m_a$, so the decay rate is only proportional to $m_a^{-2}$ instead of $m_a^{-4}$
\begin{align}\label{estimation}
\Gamma_{eff}\simeq
0.73\times 10^{-28}s^{-1}\L\f{\ld_X\ld_{C}}{0.02}\R^2\L\f{0.5\rm\,GeV}{M_{X}}\R^3
\L\f{0.1\rm\,GeV}{M_S}\R^2\L\f{N_C^2Q_C^4}{100}\R
\L\f{100\rm\,GeV}{m_C}\R^2,
  \end{align}
where $\delta=7.0$ keV has been fixed.

Considering DM relic density, we impose further conditions on the parameters. Here DM is a mixture of cold DM and WDM, the WIMP-system and gravitino, so the small scale problem is solved. But as usual the thermal gravitino would over close the Universe if there were no late entropy release, of order 100. The charged particles can provide such
source of extra entropy. This can be achieved by introducing small mixings, given $C/{\bar C}$ carrying proper SM charges, between $C/{\bar C}$ and the SM matters, which open the decay channels of $C/{\bar C}$. However, $X_R$ is also diluted, so we should suppress its annihilation rate to some degree. If $M_X>M_S$, then the typical annihilation cross section of $XX^*\ra S^*S$ scales as
\begin{align}\label{}
\sigma v\sim \f{\ld_X^4}{64\pi}\f{1}{ M_{X}^2}.
 \end{align}
Thus, one needs a relatively small $\ld_X\sim10^{-2}$ (for $M_X\sim1$ GeV) to ensure that its relic density before dilution is about $\Omega_{X}h^2\sim {\cal O}(0.1{\cal S})$. Alternatively, one can consider a proper degeneracy between $M_X$ and $M_S$ or even the non-thermal production to produce correct relic density of $X_R$.

The working model contains quite a few scales, so it is regarded as an effective model only. There is a way to update it, which is based on a single-scale dark sector
\begin{align}\label{}
W=&\f{\ld_X}{2}X^2S+\L {\cal F}S-\f{\kappa}{3}S^3\R+\ld_C SC\bar C.
  \end{align}
Here, ${\cal F}$ is a dimension-two parameter which may be dynamically generated by further embedding the model into a super QCD sector, where $X$ is identified with a meson. We leave this interesting possibility for future
 specifical study. The $F-$flantess of $S$ forces it to develop a VEV, $\langle S\rangle=\sqrt{{\cal F}/\kappa}$. In fact, $F_S=0$ is also necessary to avoid its dangerous contribution to a large mass splitting between $X_{R}$ and $X_I$. Now employing a shift $S\ra S+\sqrt{{\cal F}/\kappa}$, the linear term is eliminated,  and all the
other particles in Eq.~(\ref{portal}) acquire masses
\begin{align}\label{}
M_X=\f{\ld_X}{\sqrt{\kappa}} \sqrt{{\cal F}},\quad  m_C=\f{\ld_C}{\sqrt{\kappa}}\sqrt{{\cal F}},\quad M_S=2\sqrt{\kappa {\cal F}}.
  \end{align}
Substituting them into $\Gamma_{eff}$, it is found that to produce the correct effective decay rate, the massive scale $\sqrt{{\cal F}}$ is at most a few GeVs, as incurs a tension between $\Gamma_{eff}$ and $m_C\sim 100$~GeV
 in this single scale model.

To end up this Section, we would like to make a brief comment on the models with a light PGSB. It can be introduced in many ways. For instance, the more complicated dark sector experiences a Peccei-Quinn (PQ) like global symmetry breaking, which generates not only the dark matter mass scale but also a GSB. It acquires mass only via anomaly and thus very light. The $Z_3-$version of the next-to minimal supersymmetric SM (NMSSM)~\cite{Ellwanger:2009dp} which possesses a superpotential $\ld NH_uH_d+\kappa N^3/3$ deserves special attention, since it naturally presents a PGSB in the PQ-limit $\kappa\ra 0$. We respectively identify the singlet $S$ and charged particles $(C,\bar C)$ with $N$ and Higgs doublets, whose mass is $m_C=\mu=\ld\langle S\rangle$. To control the term $\ld_X \ld v_uv_d/2$ which leads to $X_{R/I}$ mass splitting, $\ld_X \ld \lesssim{\cal O}(10^{-5})$ or even smaller is required. However, in the case of low scale SUSY breaking, it is unclear whether or not the soft terms of the NMSSM successfully produces $\langle S\rangle$ and other acceptable phenomenological consequences.

\section{The keV Axino Models}

A light PQ axion is the most elegant solution to the strong CP problem and its supersymmetric version predicts a fermonic superpartner, the axino $\wt a$. This particle can be a good dark matter candidate~\cite{Covi:2009pq}, with mass varying in a wide range depending on the relative magnitude of SUSY-breaking and PQ-breaking scales, and the mediation mechanism to the PQ sector. So its mass generically is assumed to be a free parameter, e.g., in the keV region interested by the 3.5 keV $X-$ray line~\cite{Kong:2014gea}. Moreover, there is an anomaly induced coupling among axino, bino, and electroweak vector bosons
\begin{align}\label{}
{\cal L}=i\f{\alpha_Y C_{aYY}}{16\pi f_a}\bar{\wt a}\gamma_5[\gamma^\mu,\gamma^\nu]\wt B B_{\mu\nu},
  \end{align}
where $f_a$ denotes the PQ-symmetry breaking scale and $C_{aYY}$ is a constant specified by the model. Then, along with the mixing between bino and active neutrino, which again is a result of the bilinear $R-$parity violating terms given in Eq.~(\ref{BRP}), it two-body decays into neutrino plus gamma, explaining the $X-$ray line~\cite{Kong:2014gea}.

We now turn back to the axino mass. In the context of global SUSY breaking, it is argued in Ref.~\cite{global:axion} that the axino mass is at the scale $m_{\wt G}^2/f_a$. However, in the context of supergravity, Ref.~\cite{SG:axino} pointed out that generical $m_{\wt a}$ is at the order of $m_{\wt G}$ although  the global SUSY result may be realized
 under the special circumstance. In the sense of generality, the low scale SUSY-breaking with keV gravitino is thus strongly favored by the decaying axino warm DM.

\section{Remarks on Dark Radiation}

The keV scale dark gaugino $\widetilde{X}$
 can explain dark radiation as well. The effective number of neutrino
species $N_{\rm eff}$ can be decomposed as $N_{\rm eff}=N_{\rm eff,SM}+\Delta N_{\rm eff}$,
where the SM value is $N_{\rm eff,SM}=3.046$.
The present observational range for $\Delta N_{\rm eff}$ from Planck+WMAP9+ACT+SPT+BAO+HST at $2\sigma$ is
$\Delta N_{\rm eff} = 0.48^{+0.48}_{-0.45}$~\cite{Ade:2013zuv}. For example, suppose that
dark gauginos are out of thermal equilibrium when $kT$ is still above the masses of muon and
electron while below all the other SM particle masses except the active neutrinos,
 we obtain
\begin{align}\label{}
\Delta N_{\rm eff} = \frac{7}{8}\left(\frac{43}{57}\right)^{4/3} \sim 0.60~,~\,
  \end{align}
which is within $1\sigma$ region. Of course, to realize the observed DM relic density,
we  need an entropy release factor ${\cal S}\sim$1000 to dilute them from Eq.~(\ref{DM-Density}).

\section{Discussions and Conclusion}

We pointed out that the 3.5 keV $X-$ray line can be explained naturally in the SSMs
with low-scale SUSY breaking. And we provided three concrete models
\begin{itemize}
  \item A keV scale dark gaugino with mass from gravity mediation
plays the role of sterile neutrino in the bilinear $R-$partiy violation scenario. So it is a decaying WDM
candidate.

  \item The compressed cold DM states in a highly supersymmetric dark sector
emit such a line by the heavier one decay into the lighter one plus two photons.
The mass splitting is related to gravitino mass by gravity mediation.

  \item Light axino decays to neutrino plus gamma in the $R-$parity violating scenario. Its mass again
 is around the gravitino mass which hints the low scale SUSY breaking.

\end{itemize}

The compressed WIMP-like DM scenario also opens new window for DM model building.  We realized the degeneracy
by the holomorphic property of superpotential
 and gravity mediated SUSY breaking. Broadly speaking, any small breaking of $U(1)$
global symmetry can lead to similar consequence. For example, in the supersymmetric inverse seesaw models where the lepton number is highly conserved, the sneutrino is a good candidate~\cite{Guo:2013sna}. Additionally, in terms of our general analyses, such models favor a lighter DM, say a few GeVs. Thereby, such features hint an interesting connection with the asymmetric DM and we leave it as an open question for future explorations.

\section*{Note added}

After the completion of this work, we noticed
that~\cite{Kolda:2014ppa} appeared on the arxiv. This paper also proposed that the light dark gaugino in the context of bilinear $R-$parity violation can produce the 3.5 keV $X-$ray line. The difference is that they considered large kinetic mixing while we considered the small kinetic mixing.

\section*{Acknowledgements}

This research was supported in part by the China Postdoctoral Science Foundation under
grant number 2013M530006 (KZ), by National Research Foundation of Korea (NRF) Research Grant 2012R1A2A1A01006053, and by SRC program of NRF funded by MEST (20120001176)
through Korea Neutrino Research Center at Seoul National University (PK), and by the Natural Science
Foundation of China under grant numbers 10821504, 11075194, 11135003, and 11275246,  and by the National
Basic Research Program of China (973 Program) under grant number 2010CB833000 (TL).

\appendix

\section{Light dark gaugino from heavy dark sector}\label{split}

In this appendix we show how a $U(1)_X$ sector with relatively heavier matters and dark gauge boson allows a much lighter dark gaugino. Quite generically, there is a pair of dark Higgs $(H,\bar H)$ developing VEVs $(v,\bar v)$ to break $U(1)_X$. At the same time, they have a large mass term $\mu_HH\bar H$ with $\mu_H\gg \sqrt{2} g_X(v^2+{\bar v}^2)^{1/2}\equiv m_X\gg {\rm keV}$, the dark gauge boson mass. Then, the dark neutralino mass matrix, in the basis $(\wt H,\wt {\bar H},\wt X)$, takes a form of
\begin{eqnarray}
M_X=\L\begin{array}{ccc}
 0 & \mu_H & m_X \sin\theta_X \\
 \mu_H & 0 & -m_X \cos\theta_X  \\
  m_X \sin\theta_X & -m_X \cos\theta_X & 0
\end{array}\R,
\end{eqnarray}
where the negative sign means that $H$ and $\bar H$ carry opposite $U(1)_X$ charges. We have defined $\tan\theta_X$ as the ratio $\bar v/ v$. For the aforementioned mass hierarchy, it is not difficult to work out the dark gaugino mass
\begin{align}
M_{\wt X}\approx m_X^2\sin2\theta_X/\mu_H.
\end{align}
As one can see, dark gaugino mass can be much smaller than $m_X$ and $\mu_H$, especially in the $\theta_X\ra 0$ or $\pi/2$ limit where a $U(1)$ fermion number arises.

\section{Compressed WIMP three-body decay}\label{three-body}

In this appendix we give the detailed calculations of the heavier state $X_R$ three-body decay into the lighter state $X_I$ plus a pair of gammas via a CP-odd boson $a$ which couples to a pair of fermions $(C,\bar C)$ with electric charge $Q_C$. The Lagrangian is given by $\mathcal{L}=- \mu_a  X_R X_I a + i \ld_C  a \bar{C} \gamma^5
C$. The amplitude square of the radiative decay process $X_R\ra X_I\gamma\gamma$ is
\begin{eqnarray}
| \mathcal{M} |^2 = \frac{ \mu_a^2 }{(Q^2 - m_a^2)^2 + m_a^2 \Gamma_a^2}
| \mathcal{M}(a \rightarrow \gamma \gamma)  |^2.
\end{eqnarray}
It is convenient to decompose the three-body phase space as
\begin{eqnarray}
\mathrm{d}\Pi_3 (p_1 \rightarrow p_2 p_4 p_5) = \mathrm{d}\Pi_2 (p_1 \rightarrow p_2 Q)
\frac{\mathrm{d}Q^2}{2\pi} \mathrm{d}\Pi_2 (Q \rightarrow p_4 p_5)~,~
\end{eqnarray}
where $p_1, p_2, Q, p_4$ and $p_5$ are momentum of particles $X_R, X_I, a$ and the two photons, respectively. Then decay width of $X_R$ is
\begin{eqnarray} \label{decaywidth}
\Gamma(X_R) 
&=&\frac{1}{2m_{X_R}} \int
\frac{ \mu_a^2 }{(Q^2 - m_a^2)^2 + m_a^2 \Gamma_a^2}
\mathrm{d}\Pi_2 (p_1 \rightarrow p_2 Q)
\frac{\mathrm{d}Q^2}{2\pi}
2 \sqrt{Q^2} \Gamma(a \rightarrow \gamma \gamma)~.
\end{eqnarray}
In the above equation, the first two-body phase space can be integrated as follows
\begin{eqnarray}
\int \mathrm{d}\Pi_2 (p_1 \rightarrow p_2 Q) = \frac{1}{4\pi} \frac{|\vec{p_2}|}{m_{X_R}}~,~\,
\end{eqnarray}
where $|\vec{p_2}| = \frac{\lambda^{1/2}(m_{X_R}^2,m_{X_I}^2,Q^2)}{2m_{X_R}}$ and the function
$\lambda(x,y,z)$ is defined as $\lambda(x,y,z) = (x-y-z)^2 - 4yz$. Also,
 $\Gamma(a \rightarrow \gamma\gamma)$ is given by
\begin{eqnarray}
\Gamma(a \rightarrow \gamma\gamma) =
\frac{\sqrt[3/2]{Q^2}\alpha^2 Q_{C}^4}{256\pi^3m_{C}^2}\ld_C ^2
| A(Q^2/4m_{C}^2)|^2~,
\end{eqnarray}
with the function $A(\tau) = 2 \tau^{-1} f(\tau)$ where
\begin{eqnarray}
f(\tau)=\left\{
\begin{array}{ll}  \displaystyle
\arcsin^2\sqrt{\tau} & \tau\leq 1 \\
\displaystyle -\frac{1}{4}\left[ \log\frac{1+\sqrt{1-\tau^{-1}}}
{1-\sqrt{1-\tau^{-1}}}-i\pi \right]^2 \hspace{0.5cm} & \tau>1 
\end{array} \right. ~.
\label{eq:ftau}
\end{eqnarray}
Therefore, we get the partial decay width of $X_R$
\begin{eqnarray} \label{simdecaywidth}
\Gamma(X_R) = \int \frac{\mu_a^2 }{64\pi^5} \alpha^2 Q_{C}^4 \ld_C ^2
\frac{\lambda^{1/2}(m_{X_R}^2,m_{X_I}^2,Q^2)m_{C}^2}{m_{X_R}^3(Q^2-m_a^2)^2}[\arctan(\frac{1}{\sqrt{\frac{4m_{C}^2}{Q^2}-1}})]^4 \mathrm{d}Q^2,
\end{eqnarray}
where the condition $Q^2/4m_{C}^2 \le 1$ is used and the decay width of $a$ is neglected. The range of $Q^2$ is
 $0 \leq Q^2 \leq (m_{X_R}-m_{X_I})^2$.

In the kinematic region under consideration, a simple analytic expression is available. Recalling that $\delta = m_{X_R} -m_{X_I} \sim$ keV, thus one has $\delta \ll m_{X_R}, m_{X_I}, m_a$ and $m_{C}$. Then we get the approximations
\begin{eqnarray} \label{lambdapro}
\lambda^{1/2}(m_{X_R}^2,m_{X_I}^2,Q^2)
&\simeq & 2m_{X_R} \sqrt{ \delta^2- Q^2},\quad \frac{1}{(m_a^2 - Q^2)^2}  \simeq m_a^4(1 + \frac{Q^2}{m_a^2})^2,\cr
\arctan^4(\frac{1}{\sqrt{\frac{4m_{C}^2}{Q^2}-1}})& \simeq&
\frac{1}{(\frac{4m_{C}^2}{Q^2}-1)^2} \simeq \frac{(Q^2)^2}{16m_{C}^4}
(1+\frac{Q^2}{4m_{C}^2})^2.
\end{eqnarray}
Substituting Eq. (\ref{lambdapro}) into Eq. (\ref{simdecaywidth}), we get
\begin{eqnarray}
\Gamma(X_R)
&\simeq &  \frac{\mu_a^2 }{3360\pi^5}  \alpha^2 Q_{C}^4 \ld_C ^2
\frac{\delta^7}{m_{X_R}^2 m_a^4m_C^2}.
\end{eqnarray}

\end{document}